\documentclass{iau} 
\usepackage{graphicx}

\usepackage{natbib}

\newcommand\aap{\textit{A\&A}}

\newcommand\mnras{\textit{MNRAS}}

\newcommand\apjl{\textit{ApJ}}

\newcommand\aj{\textit{AJ}}

\newcommand\kms{{\rm\,km\,s^{-1}}}

\title[Kinematic structures found with Gaia DR1/TGAS and RAVE in the Solar neighbourhood] 
{Kinematic structures found with Gaia DR1/TGAS and RAVE data in the Solar neighbourhood}

\author[I.~Kushniruk et al.]   
{I.~Kushniruk$^1$, T.~Schirmer$^1$, T.~Bensby$^{1}$}

\affiliation{
$^1$ Lund Observatory, Box 43, SE-221\,00 Lund, Sweden\\
}

\pubyear{2017}
\volume{334}  
\jname{Rediscovering our Galaxy}
\editors{C. Chiappini, I. Minchev, E. Starkenburg, \& M. Valentini, eds.}
\begin{document}

\maketitle

\begin{abstract}
With the much enlarged stellar sample of 55\,831 stars and much increased precision in distances, proper motions, provided by Gaia~DR1 TGAS we have shown with the help of the wavelet analysis that the velocity distribution of stars in the Solar neighbourhood contains more kinematic structures than previously known. We detect 19 kinematic structures between scales 3-16$\kms$ at the $3\sigma$ confidence level. Among them we identified well-known groups (such as Hercules, Sirius, Coma Berenices, Pleiades, and Wolf 630). We confirmed recently detected groups (such as Antoja12 and Bobylev16). In addition we report here about a new kinematic structure at $(U,V)\approx(37, 8)\kms$. Another three new groups are tentatively detected, but require confirmation. 

\keywords{Galaxy: formation, Galaxy: evolution, Galaxy: structure, Stars: kinematics and dynamics}
\end{abstract}

\firstsection 
\section{Introduction}

Recent studies have shown that the velocity distribution of stars in the Solar neighbourhood is rich with kinematic structures, with stars that for various reasons have similar space velocities $(U,V,W)$  \cite[e.g.:][]{_dehnen98, _skuljan99, _famaey05, _antoja08, _zhao09, _antoja12}. The list of works on kinematic groups could be extended and all of them prove that the velocity distribution in the Solar neighbourhood is inhomogeneous and has a complex, branch-like structure. The question on how did the stellar streams formed is still actual. In this work we focus on the velocity distribution of stars in the $U-V$ plane for the enlarged stellar sample that is based on the astrometric data provided by Gaia DR1. 

\section{Input data and methodology}

The astrometric data provided by Gaia~DR1 (TGAS) \cite[][]{_michalik15} and was accomplished with radial velocities from RAVE~DR5 \cite[][]{_kunder17}. 
To get a better precision of positions of the structures we cut $U$ and $V$ velocity uncertainties to be less than $4\,\kms$. This limit gives us a 
sample of 55\,831 stars to which we applied a wavelet analysis with the ``a trous'' algorithm. To filter the output data we used an auto-convolution histogram method.  We then run 2\,000 Monte Carlo simulations to verify that the detected structures are real due to velocity uncertainties. 
\section{Main results}

\begin{figure}
\centering
\resizebox{0.85\hsize}{!}{
\includegraphics[viewport= 0 0 600 400,clip]{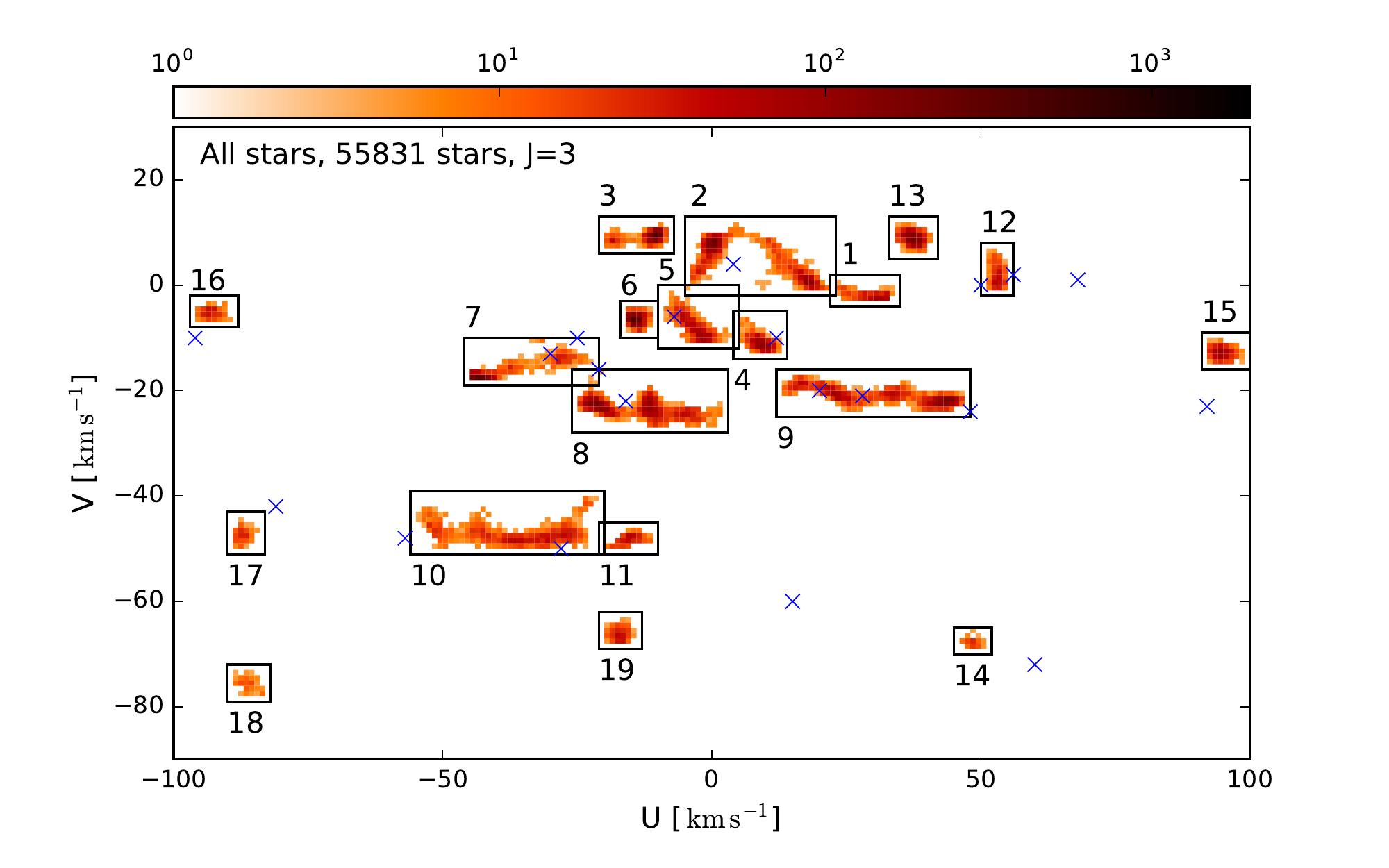}
\includegraphics[viewport= 40 40 280 412,clip]{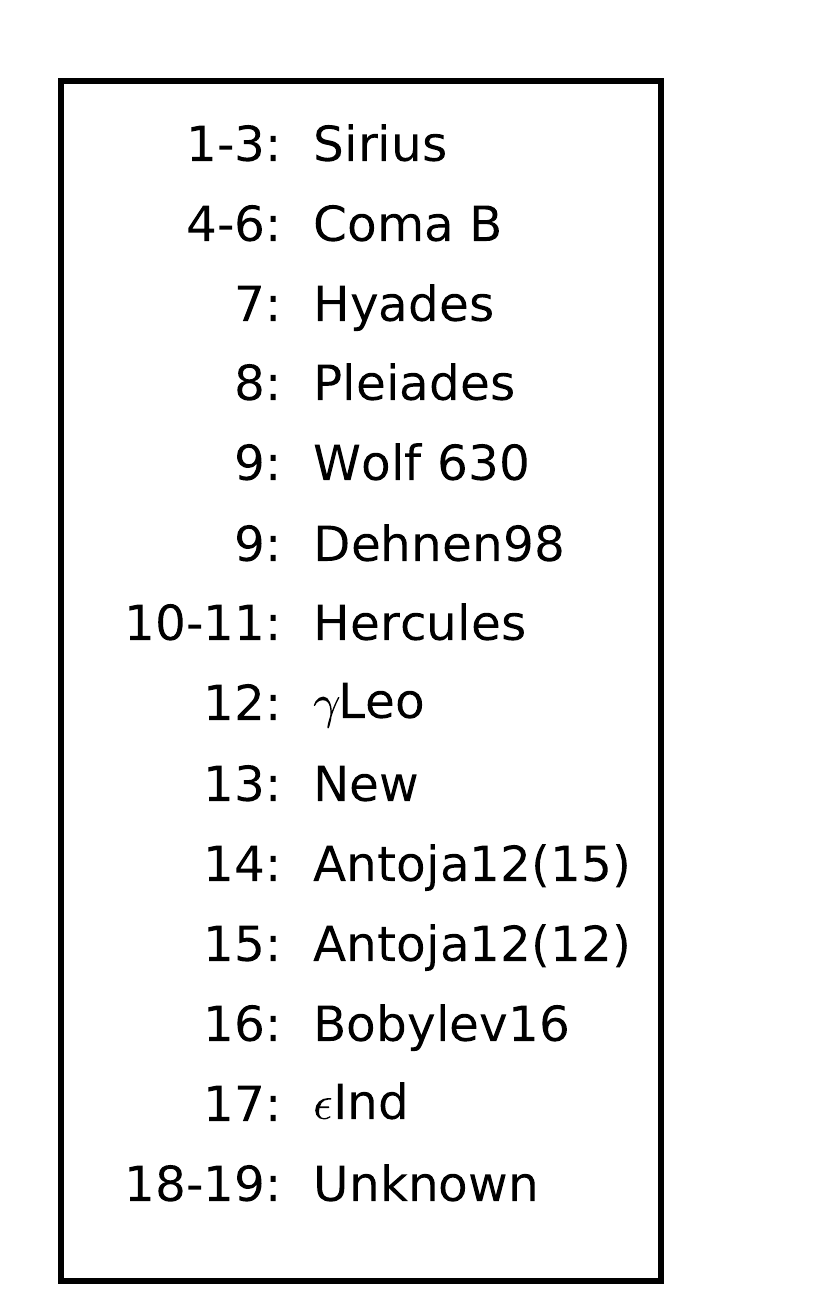}
}
\caption{Positions of kinematic stellar structures obtained by wavelet transform applied for $N_{MC}=2\,000$ synthetic data samples for $J=3$ in the $U-V$ plane. Structure counts are shown with the orange colour. Black boxes embrace region of individual structures. Blue crosses show identification of structures in literature if any. 
\label{_j3}
}
\end{figure}

Figure~\ref{_j3} shows $3\sigma$-significant (99.86~\% means that the structure is not an artefact) detected structures in the $U-V$ plane for the scale $J=3$  (structure sizes are in the range 3-16$\kms$). Here previously detected structures found in the literature (\citet{_eggen96, _antoja08, _antoja12, _bobylev16}) are marked with blue crosses. Classical moving groups like Sirius (1-3), Coma Berenices (4-6), Hyades (7), Pleiades (8) and Hercules (10-11), and some smaller structures like Wolf 630 (9), Dehnen98 (9), $\gamma$Leo (12) can be easily recognized. They all have a comparably high percentage of detection ($> 50\%$ of repeats in Monte Carlo simulations) and a big number of stars ($>$ 100). We confirm the two structures from \cite{_antoja12} (structures 14 and 15) and one structure from \citet{_bobylev16} (structure 16). A new structure (number 13) was detected with 74\% of significance. Groups 18-19 have low percentages of detection: $<15\%$, and might be insignificant. For more details see \citet{_kushniruk17}.

Together in a combination with the final RAVE~DR5 data we confirm results obtained before the Gaia era that the velocity distribution of stars in the solar neighbourhood is inhomogeneous and completed the list of known structures with a few more new groups. 

\begin{acknowledgments}
T.B. was funded by the ``The New Milky Way'' project grant from the Knut and Alice Wallenberg Foundation. We thank Prof. F.~Murtagh for making available for us the MR software packages and for valuable and helpful comments. 
\end{acknowledgments}

\end{document}